\documentclass{svproc}
\usepackage{url}
\usepackage{graphicx}
\usepackage{comment}

\usepackage{lipsum}
\usepackage{mwe}

\begin{document}
\mainmatter              

\title{
CodeAPeel: An Integrated and Layered Learning Technology For Computer Architecture Courses}
\titlerunning{CodeAPeel-C}  
%

\author{

\vspace{-15pt}

A. Yavuz Oruc\inst{1},  Abdullah Atmaca\inst{2}, 
Y. Nevzat Sengun\inst{3}, A. Semi Yenimol\inst{3} }
\authorrunning{A. Yavuz Oruc et al.} 
%
\tocauthor{A. Yavuz Oruc, Abdullah Atmaca, Y. Nevzat Sengun, and A. Semi Yenimol}

\institute{University of Maryland, College Park, MD 20740, USA,\\
\and
Everon, Amsterdam, Netherlands
\and
Bilkent University, Ankara, Turkey}

\maketitle              

\vspace{-25pt}
\begin{abstract}
This paper introduces a versatile, multi-layered technology to help support teaching and learning core computer architecture concepts. This technology, called CodeAPeel is already implemented  in one particular form to describe instruction processing in compiler, assembly, and machine layers of a generic instruction set architecture by a comprehensive simulation of its fetch-decode-execute cycle as well as  animation of  the behavior of its CPU registers, RAM, VRAM, STACK memories, various control registers, and graphics screen.  Unlike most educational CPU simulators that simulate a real processor such as MIPS or RISC-V, CodeAPeel  is designed and implemented as a generic RISC instruction set architecture simulator with both scalar and vector instructions to provide a dual-mode processor simulator as described by Flynn's classification of SISD and SIMD processors. Vectorization of operations is built into the instruction repertoire of CodeAPeel, making it straightforward to simulate such processors with powerful vector instructions.
\vspace{-5pt}
\end{abstract}
\vspace{-30pt}
\section{Introduction}

\vspace{-5pt}\noindent
Computer organization and architecture courses are  taught  regularly in most electrical and computer engineering as well as computer science curricula in universities and colleges across the globe.  It is common practice to use software tools to teach key concepts in such courses. This practice often relies on simulation tools to demonstrate the instruction fetch, decode, and execution process in processor chips by providing a view of how  such a process affects various parts of a computer, including the control unit,  instruction register, program counter,  datapath and operand registers, cache,  core memory, and peripheral devices. Simulation and modeling tools that describe computer operations have been around nearly as long as computers existed.  The earliest  computer architecture simulators were introduced to model the behavior of various subsystems within a computer system during 1960s and 70s, when computers were programmed using punch cards \cite{earlySurvey,earliestCPUSimulator,secondearliestCPUSimulator,assemblyLanguageCourse,interactiveVisualSimulator,thirdearliestCPUSimulator,simulatorOnUnivac,HPCalAssemblyProgramming,instructionalSimulator,fourthearliestCPUSimulator}. Those in \cite{secondearliestCPUSimulator,assemblyLanguageCourse,interactiveVisualSimulator,thirdearliestCPUSimulator,simulatorOnUnivac,HPCalAssemblyProgramming} were particularly developed for educational purposes and run on IBM S/360, CDC 6500, CDC 6600, Univac 1110 computers, and an HP-54 calculator.  The simulators presented in\cite{assemblyLanguageCourse,instructionalSimulator} were designed to describe hypothetical computer architectures. With the arrival of multiprocessor computer systems in late 1970s  and early 80s, a number of multiprocessor simulators were introduced to describe the interactions between  processors  in such computer systems\cite{connectionistSimulator,messagePassingMultiprocessor,butlerOrucICDCS,butlerOrucMicroJ,butlerOrucPBS,multiprocSimulator,parallelSimulatorChip}. While these multiprocessor simulators have been useful for computer architecture research, they are not as suitable as educational simulation tools for core computer architecture courses, especially at the undergraduate level. 
During the last three decades,  several other software tools were introduced for complete as well as  partial  machine simulations of real CPUs and hypothetical processors. Such tools include Shade\cite{Shade1994}, SMT\cite{Tullsen1996}, SIMOS\cite{simOS},  SimpleScalar\cite{SimpleScalar2002}, Simics\cite{simics2002},  PTLSim\cite{ptlSIM},  Multi2Sim\cite{Multi2Sim2007,Multi2Sim}, GEM5\cite{GEM5}, and MARSSx86\cite{marssx86}. Some of these tools were developed with cycle and/or time accurate simulation capabilities\cite{survey2019}. As such,  they are  valuable for validating and verifying architectural specifications and system requirements  during the design, development, and testing stages of commercial processor chips such as MIPS, ARM, ALPHA, SPARC,  RISC,  and Intel x86. However, they are overly complex to set up and run, especially for simple simulation tasks such as running an assembly language or machine program on a simulated CPU. Often they need to be compiled from source files with a multitude of auxiliary programs and libraries,  and  require  command line processing before they can be used. Moreover, the simulation process typically involves compiling and executing programs written in languages such as C and Python,  and simulation results are dumped to text files. These issues make most of these tools unattractive for a dynamic and interactive computer architecture instruction. 
Other simulation tools were developed  specifically as hands-on instructional software for computer organization courses. Examples of such instructional tools include SPIM\cite{larus1990}, 
EasyCPU, Little Man Computer, RTLSim\cite{easyCPU2001}, Marie\cite{Marie2003},   MARS\cite{MARS2006},  EDUMIPS64\cite{eduMips,eduMIPS64}, and DEED\cite{deed2014,simulators2016}. Even though these simulators are more educationally-driven and can be run as an ordinary computer application, they have limited simulation capabilities as they do not allow modifying register bit-width, scratchpad or memory sizes, instruction types, and operand bit-widths. They simulate one particular processor or another, focusing mostly on the  timing of pipeline executions of machine programs, overlooking the interactions between the CPU and peripheral devices.   Moreover, their user interfaces lack depth and detail, limiting their utility as an effective teaching or learning tool as described in the next section.

In this paper, we present a comprehensive approach to the design and implementation of a learning technology that provides a layered view of computers from the compiler down to machine language executions of computer programs. The framework of this approach is referred to as CodeAPeel. This framework supports a load/store instruction set architecture whose instruction repertoire  includes both scalar and vector instructions. Vector processing has become a natural method of extending the degree of parallelism in contemporary processor chips, with RISC-V and other RISC architectures providing SIMD extensions in their instruction repertoires\cite{armVector,vectorprocessingAsanovic,vectorization}. The initial ideas of CodeAPeel were originally implemented in a more rudimentary instruction set architecture and software tool with scalar instructions,  called CodeMill\cite{CodeMill2003}. CodeAPeel is a much more comprehensive simulator with both functional and timing simulation features for SISD as well as SIMD computer architectures as described in\cite{Flynn66}. 

\vspace{-10pt}
\section{A Brief Survey of  Educational CPU Simulation  Tools}
\label{section:survey}

\begin{figure}
\vspace{5pt}
    \begin{minipage}{0.5\textwidth}
        \centering
        \hspace{-30pt}\includegraphics[width=0.59\textwidth]{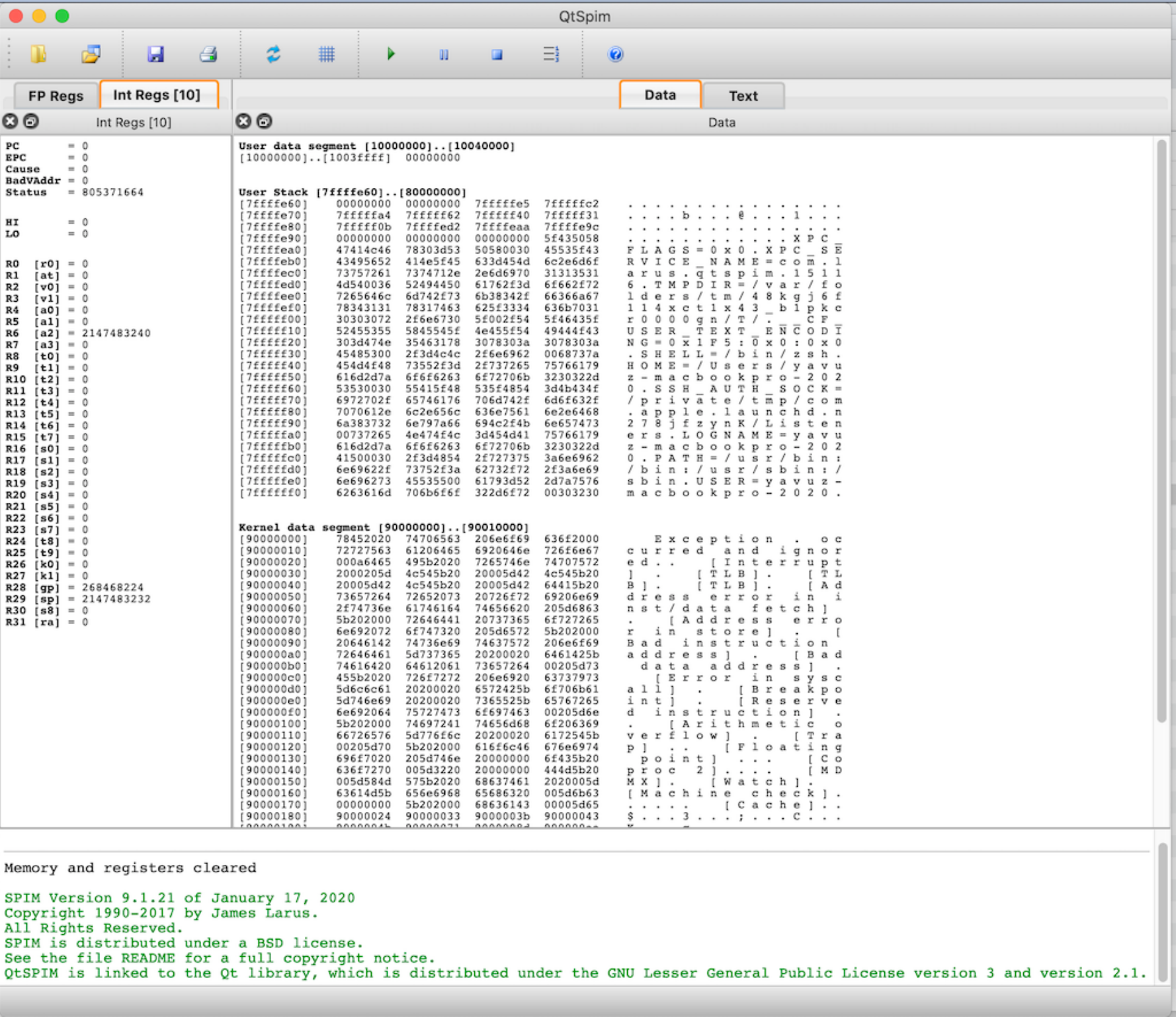} 
        \caption{QtSPIM\cite{qtSpim}.\,\,\,\,\,\,\,\,\,\,\,\,\,\,\,\,\,\,}
        \label{fQtSPIM}
    \end{minipage}\hfill
    \begin{minipage}{0.5\textwidth}
        \hspace{-30pt}\includegraphics[width=0.99\textwidth]{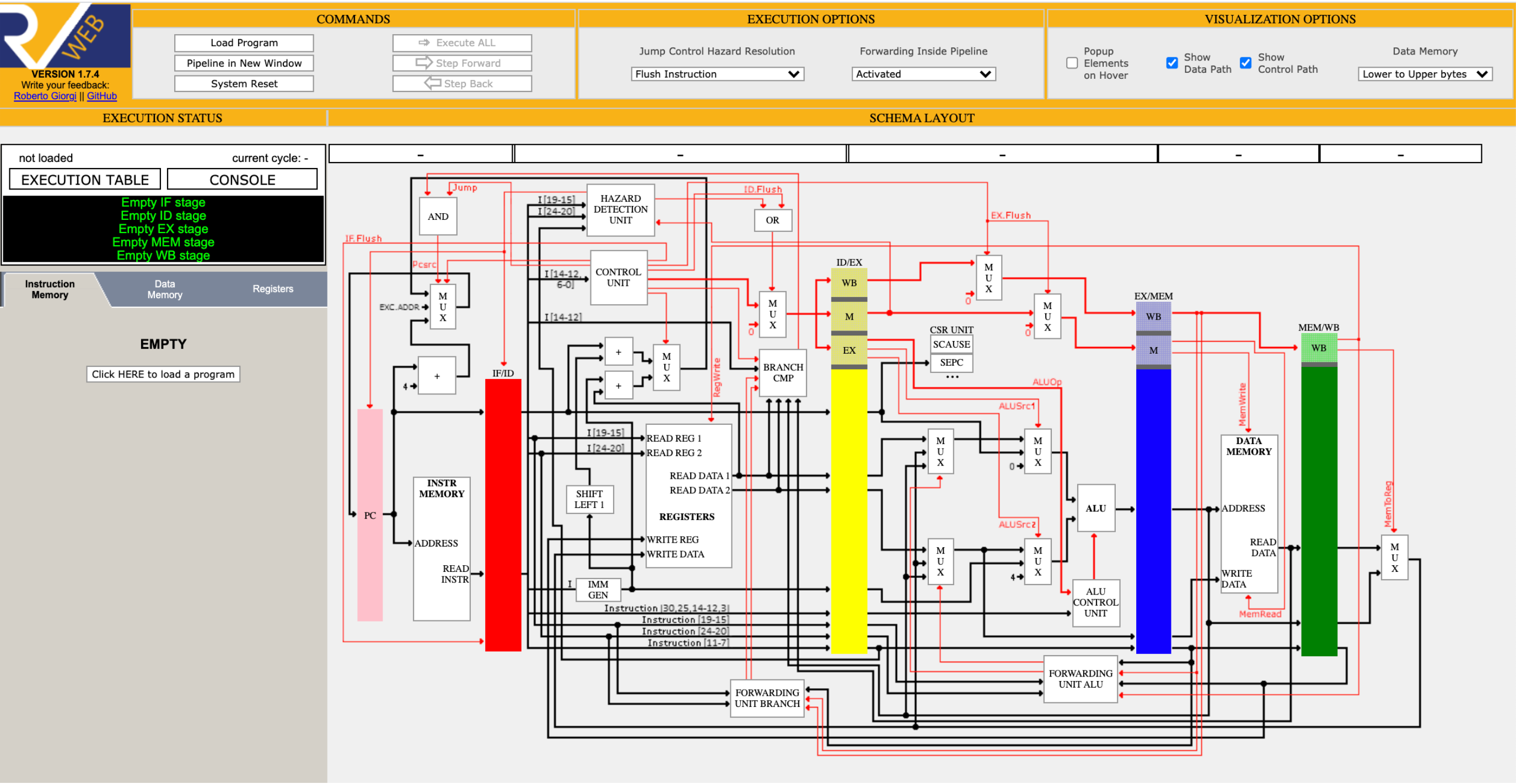} 
        \caption{WebRISC-V\cite{webRISC-V}.\,\,\,\,\,\,\,\,\,\,\,\,\,\,\,\,\,\,\,\,\,\,\,\,\,\,\,\,\,\,}
        \label{fWebRISC-V}
    \end{minipage}
\end{figure}

\begin{figure}
\vspace{-5pt}
    \begin{minipage}{0.62\textwidth}
        \centering
        \hspace{-30pt}\includegraphics[width=0.59\textwidth]{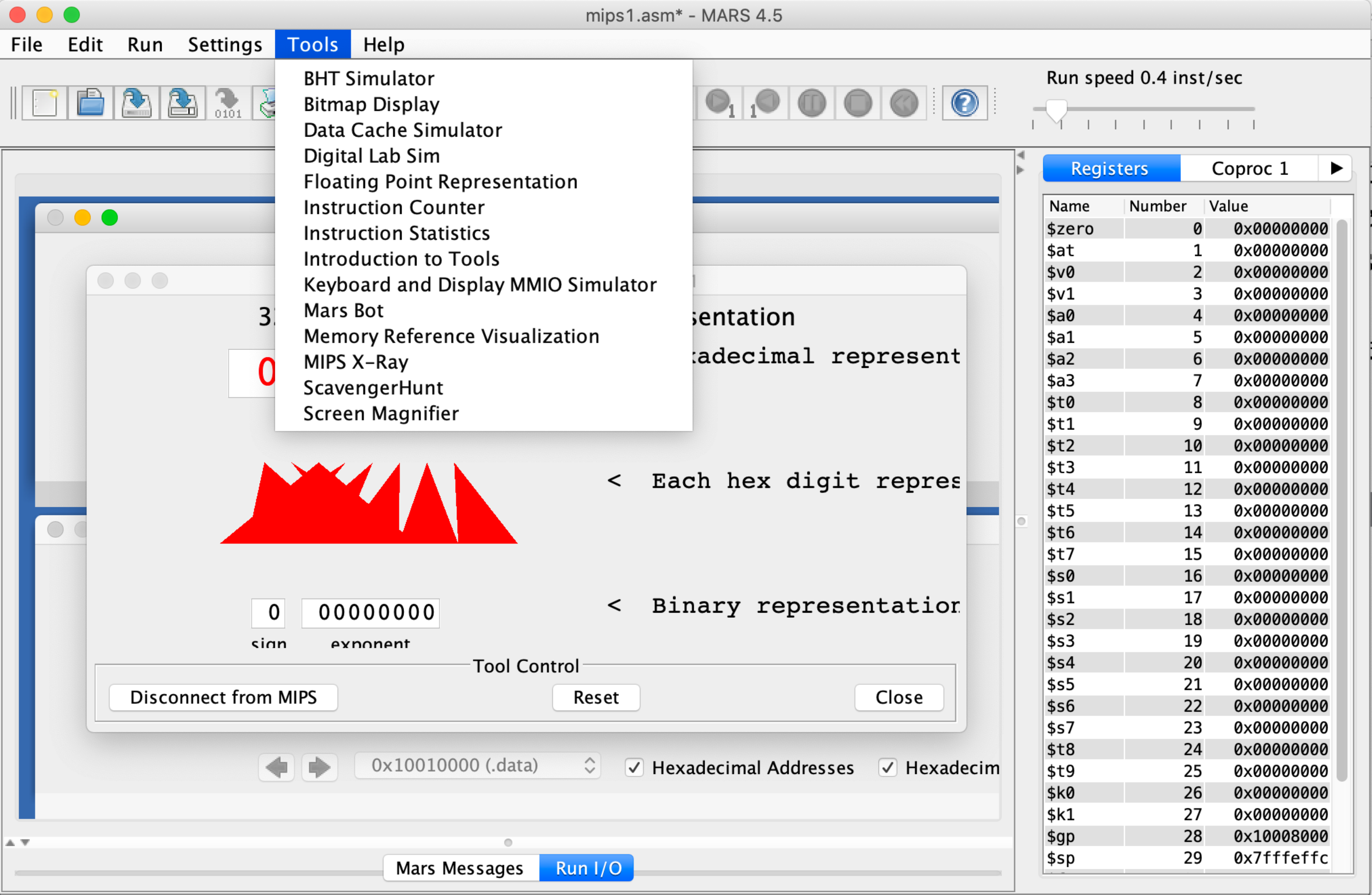} 
       \caption{MARS\cite{MARS2006}.\,\,\,\,\,\,\,\,\,\,\,\,\,\,\,\,\,\,\,\,\,\,\,}
       \label{fMARS}
    \end{minipage}\hfill
    \begin{minipage}{0.38\textwidth}
        \hspace{-30pt}\includegraphics[width=0.99\textwidth]{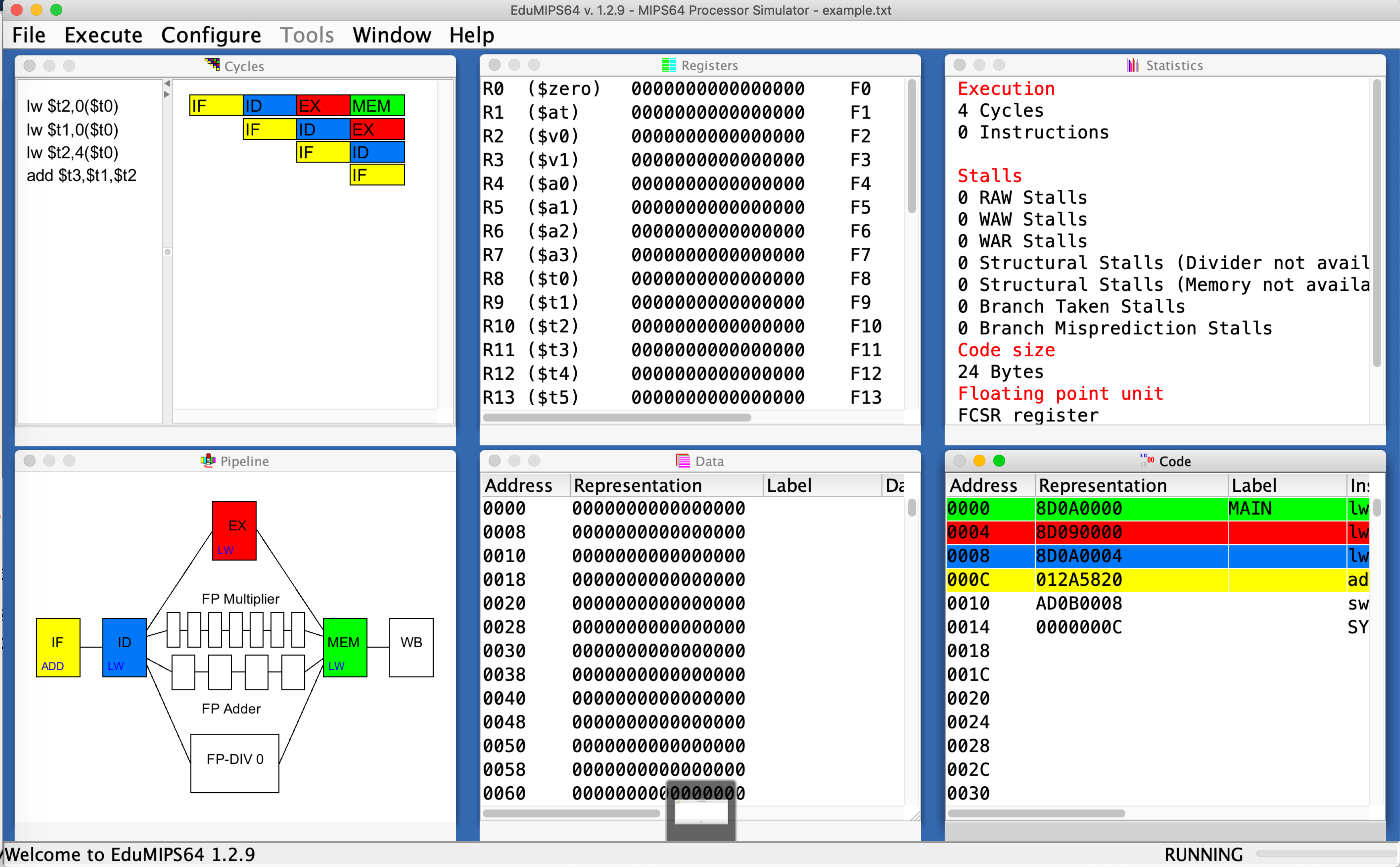} 
        \caption{EDUMIPS-64\cite{eduMIPS64}.$\,\,\,\,\,\,\,\,\,\,\,\,\,\,\,\,\,\,\,\,\,\,\,\,\,\,\,\,\,\,\,\,$}
       \label{fEDUMIPS-64}
    \end{minipage}
\end{figure}

\begin{figure}[t]
\vspace{1pt}
\centerline{\includegraphics[scale=0.26]{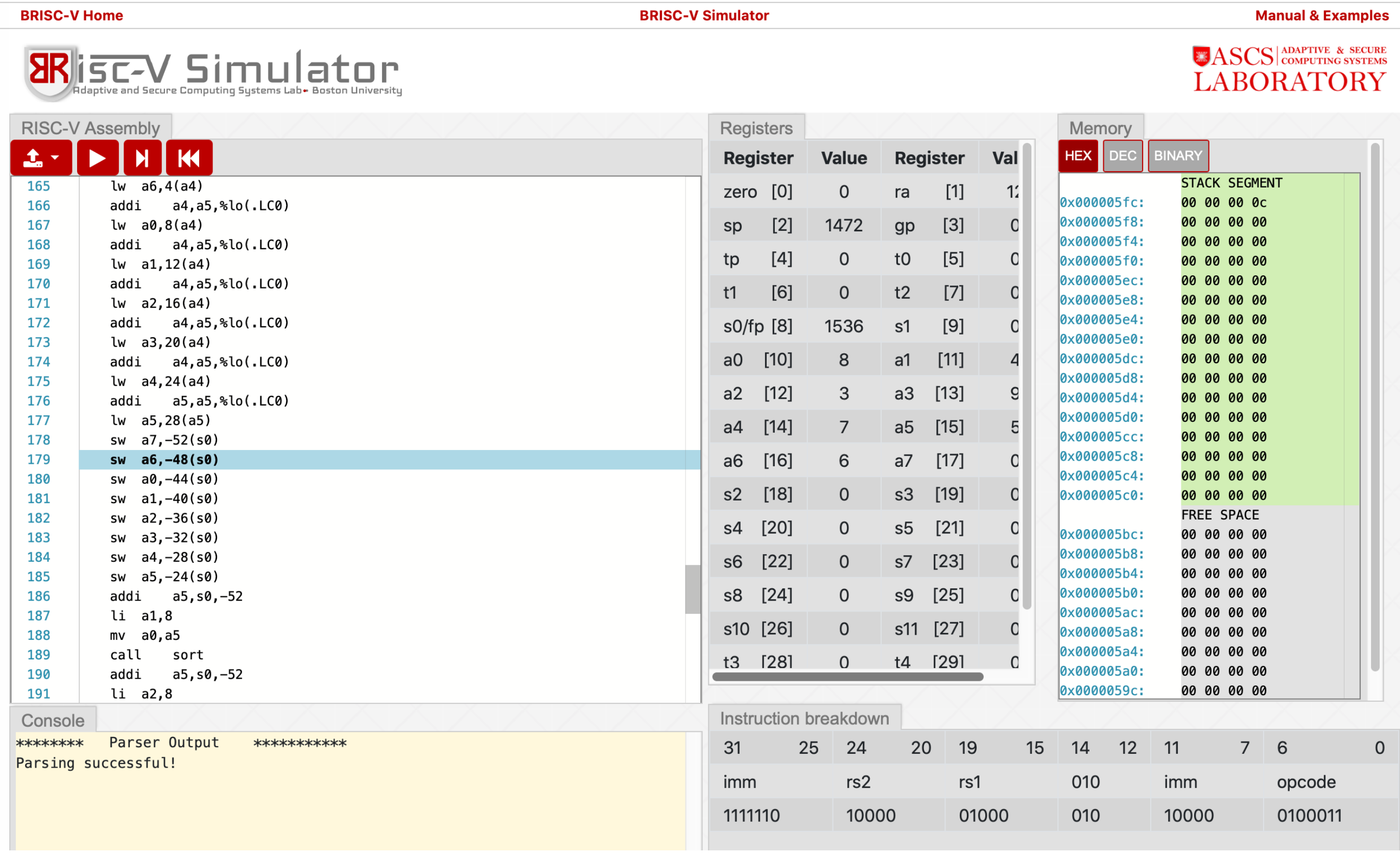}}
\vspace{-6pt}
\caption{BRISC-V\cite{brisc-v}}
\label{fBRISC-V}
\end{figure}

\vspace{-8pt}\noindent
In this section, we present a brief description of some of the CPU simulators that are used as instructional tools in computer architecture courses.  For an in-depth survey, we refer the reader to excellent articles in~\cite{survey2019,simulators2016,survey2009}. The last two references cover educational  simulators and their effectiveness, while the first one deals with  both educational and research-driven simulators.
SPIM\cite{larus1990} and its successor QtSPIM are perhaps the most-widely used machine instruction-level simulators in an educational setting, possibly due to the popularity of MIPS architecture\cite{hennessy}. SPIM\cite{larus1990} was first released in 1990 and subsequent cross-platform versions appeared periodically since then and through 2020. A history of all versions of MIPS through 2011, and their release dates can be found in\cite{qtSpim}. 
The QtSPIM\cite{qtSpim}, the current version of SPIM, running on a MacBook as of 2020 is shown in Figure~\ref{fQtSPIM}. As seen in the figure, the user interface of QtSPIM supports loading,  stepping through, and executing MIPS  machine programs. Its layout and user interface provide a simplistic representation of the execution of MIPS32 machine programs with a rudimentary  access to some operating systems functions to open files and print messages.  
WebMIPS and WebRISC-V are  alternative online simulators that are focused on pipeline executions of MIPS and RISC-V instructions\cite{webMIPS,webRISC-V}. The user interface of the latter simulator is shown in Figure~\ref{fWebRISC-V}. Both simulators have a program window that allows the user to enter and run MIPS and RISC-V machine programs. WebRISC-V is more dynamic and functional than WebMIPS. However, the animation of the instruction flow in both is too focused on buffering between pipeline stages. 
MARS is another MIPS simulation tool, which was written in Java with more extensive capabilities for stepping, running, and visualizing CPU registers and memory locations\cite{MARS2006}. 
It includes a data cache simulator, animation of MIPS datapath, branch history table, floating-point number display, single layer graphics display, a text-based output window, and keyboard simulator, among other features such as  providing  basic statistics about the type of instructions, which are executed during a sample run. Figure~\ref{fMARS} shows the layout of MARS.
EduMIPS64  is another MIPS simulator that is currently active on GitHub as an open source Java application\cite{eduMips,eduMIPS64}. It has several windows, displaying registers, and pipeline animation for a multi-function MIPS pipeline as shown in Figure~\ref{fEDUMIPS-64}.
Another web-based simulator, called BRISC-V was reported in\cite{brisc-v} for simulation of  RISC-V instructions. This simulator is capable of compiling C/C++ code into a RISC-V assembly language program and running it with stepping and execution features. Figure~\ref{fBRISC-V} shows the snapshot of BRISC-V during stepping through a binary search RISC-V assembly language program, where the window on the left displays the program and the window on the right shows how registers are updated as the program is executed. A particular edition of CodeAPeel, called CodeAPeel-C provides a similar function in its most basic view, but with many more options that include  clock speed, byte-ordering, pixel processing, input/output handling, and stack processing.
More recently, a number of other RISC-V simulators have been introduced. Spike RISC-V is one such simulator that provides a terminal-driven simulation of base RISC-V 32 and 64-bit architectures and several extensions\cite{spike}. Other RISC-V simulators have been reported in~\cite{vectorization,RISC-VEmulator,AnotherRISC-VSimulator,DrMips} with similar features. Along a different direction, software tools such as  HASE and CPU-SIM focus on the design aspects of computer architecture, and provide a design space, where instructions, operands, and various computer architecture components can be combined to create a desirable architecture for testing\cite{haseEdinburgh,cpuSim}.   As stated at the beginning of this section, we described some specific instructional  software tools in use today to illustrate the scope and style of a representative set of computer architecture educational simulators. In what follows, we present CodeAPeel-C computer architecture simulation software and compare it with some of these  simulators.

\vspace{-5pt}
The rest of the paper is organized as follows.  Section~\ref{section:codeapeelOverview} presents an overview of CodeAPeel's design goals and its roadmap. Section~\ref{section:applicationModel} describes the application model of CodeAPeel as a Java project. Section~\ref{section:assemblerDisassembler} presents assembler and disassembler functions in CodeAPeel-C.  Section~\ref{section:vectorization} explains how vectorization is implemented in CodeAPeel-C.  Section~\ref{section:screenProgramming} demonstrates how multi-layer graphics  and screen programming are integrated into CodeAPeel-C, and  Section~\ref{section:sampleRuns} demonstrates CodeAPeel-C's assembly and machine language simulation features with three sample runs. Section~\ref{section:otherTools} provides a  comparison of CodeAPeel-C with other computer architecture simulation tools and the paper is concluded in Section \ref{section:conclusion} with a discussion of  possible directions for future research. 

\vspace{-10pt}
\section{Overview of CodeAPeel-C Architecture}
\label{section:codeapeelOverview}

\vspace{-8pt}\noindent
\begin{figure}[t]
\vspace{-1pt}
\centerline{\includegraphics[scale=0.25]{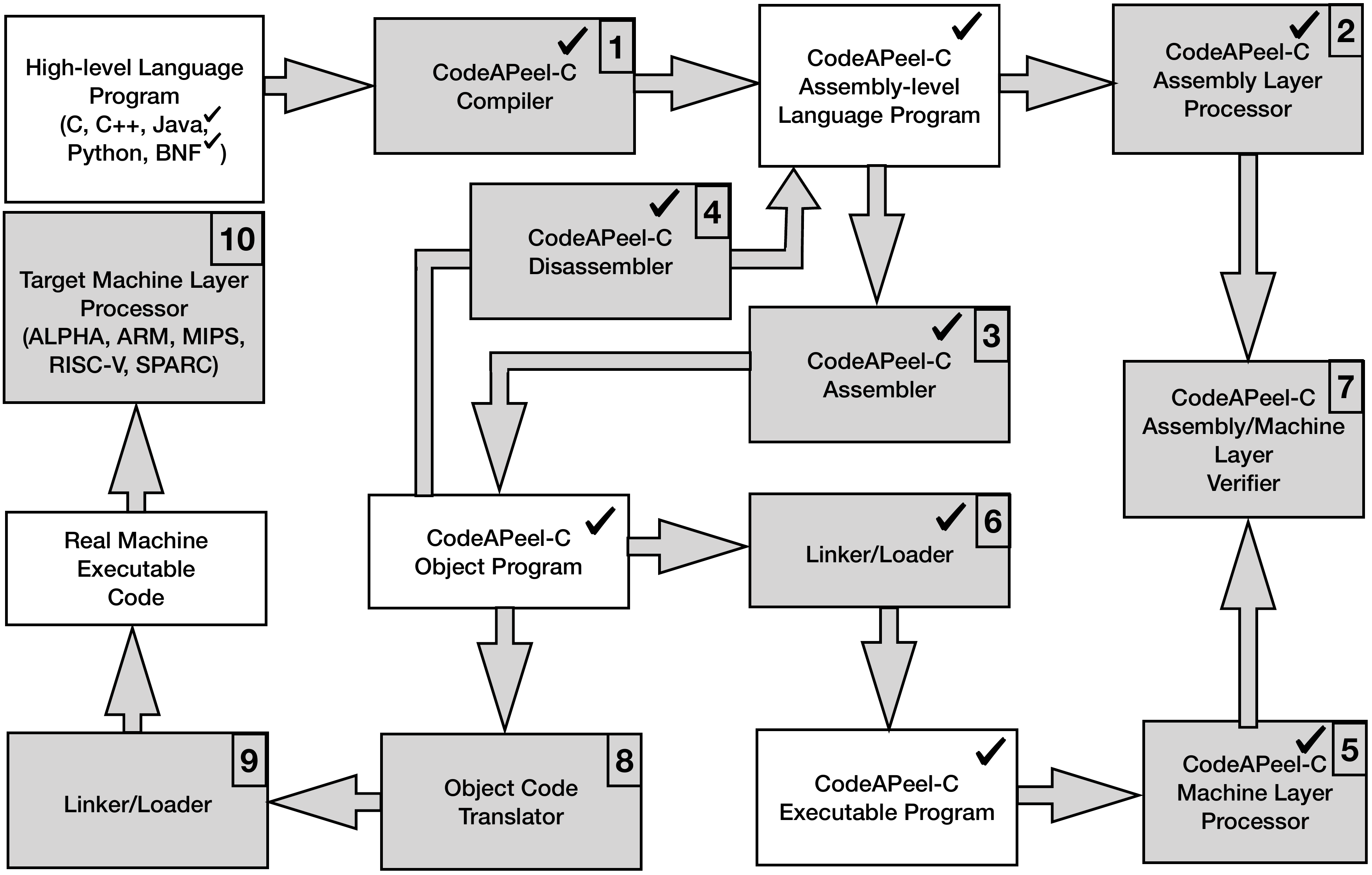}}
\vspace{-6pt}
\caption{CodeAPeel-C design map.}
\label{codeapeelOverview}
\end{figure}
CodeAPeel-C is one particular edition  within the CodeAPeel  design and development framework that was conceived to create a comprehensive learning tool  by integrating the simulation and animation of many layers of  a computer system into a single application that can accurately and realistically describe the various operations  within that system. Figure~\ref{codeapeelOverview} displays the blueprint of CodeAPeel-C. The main goal of CodeAPeel-C initiative is to make a seamless bidirectional transition between compiler, assembly language, and machine layers of computers. We have already implemented CodeAPeel-C's compiler, assembler, disassembler, assembly-level and machine language processors, and loader. Assembly and machine language CodeAPeel-C processors support the execution of 96  instructions that extend from load and store operations with several addressing modes to arithmetic, shift, logic, data transfer, branch, stack, graphics and input/output processing operations. We expect that the remaining engines (unchecked boxes in Figure~\ref{codeapeelOverview} will be operational during the next six months.  These remaining engines aim to extend the CodeAPeel-C simulation environment  in order to support executions of assembly and machine language programs on  real instruction set architectures. Thus, when completed, CodeAPeel-C will provide a comprehensive computer architecture simulator, describing  compiler, assembly language, and machine layers of a number of real architectures such as  MIPS and RISC-V as well.

\begin{figure}[t]
\vspace{8pt}
\centerline{\includegraphics[scale=0.22]{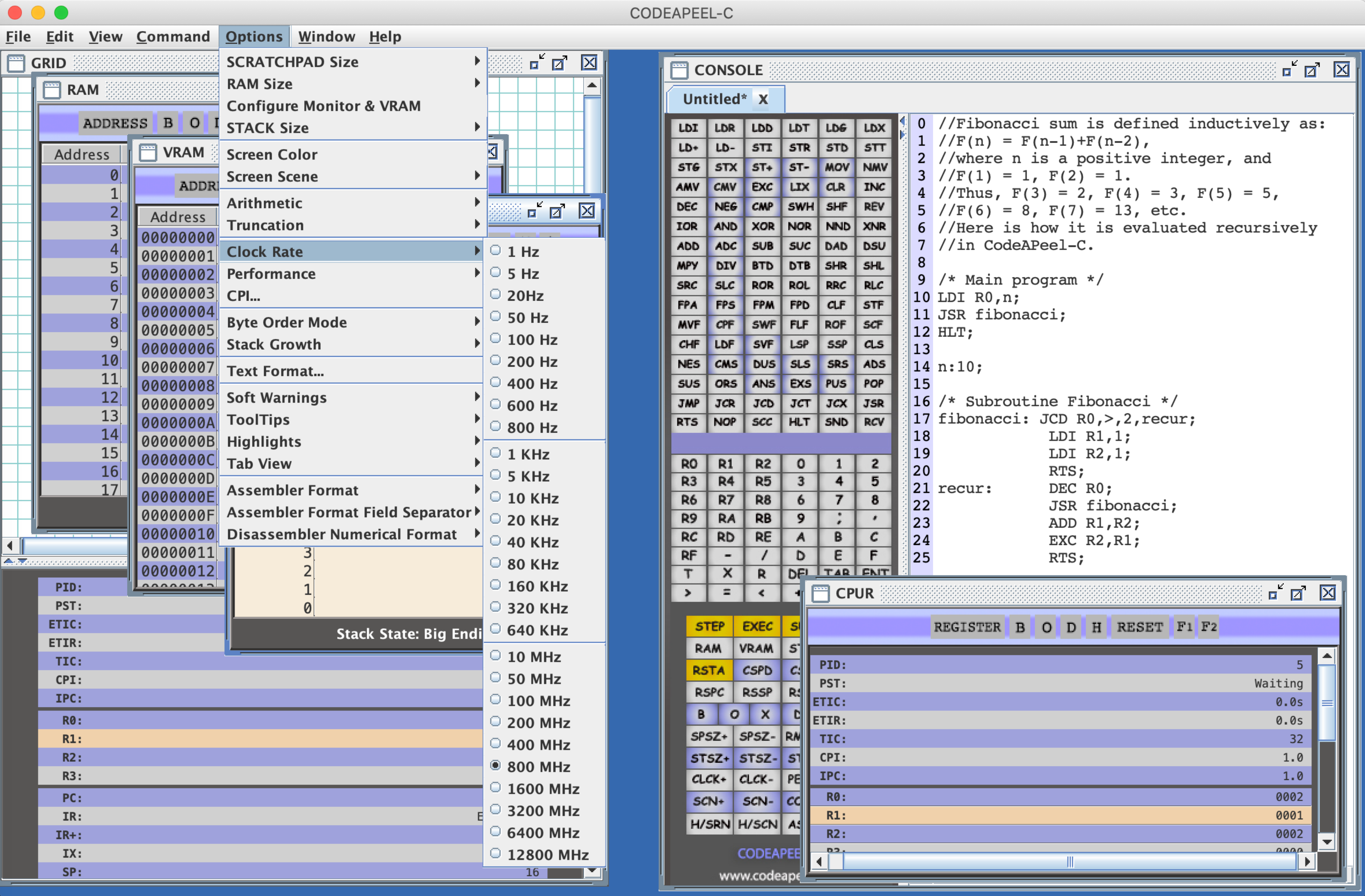}}
\vspace{2pt}
\caption{CodeAPeel-C application user interface.}
\label{userInterface}
\end{figure}

\begin{figure}[t]
\vspace{3pt}
\centerline{\includegraphics[scale=0.22]{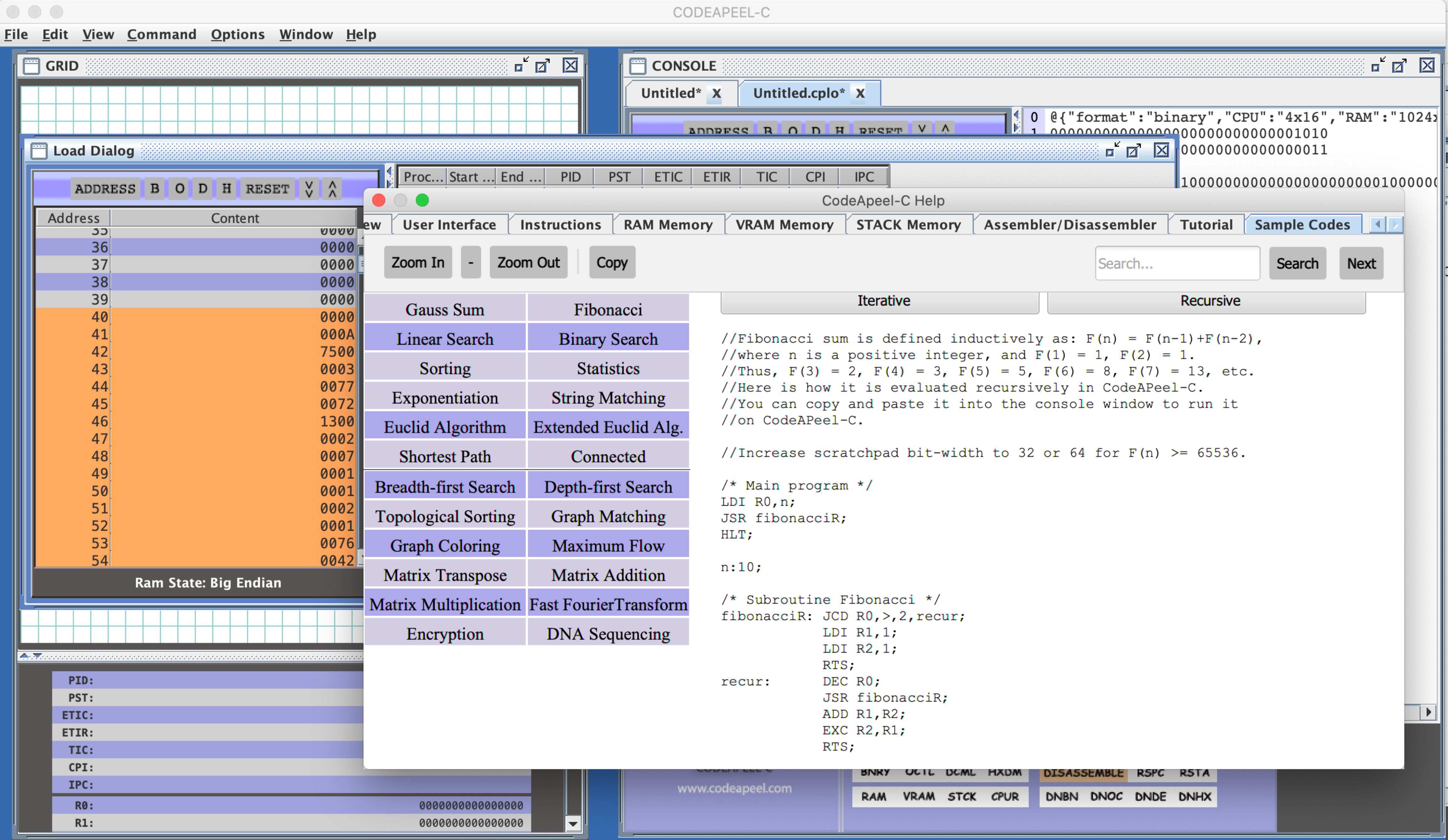}}
\caption{CodeAPeel-C help window.}
\label{helpWindow}
\end{figure}

Figure~\ref{userInterface} shows the user interface of CodeAPeel-C application. As  assembly-level language programs in the window (CONSOLE) on the right are stepped through or executed, CodeAPeel-C computer's behavior is simulated in the graphics screen and data and control registers in the window on the left, while various statistics about this behavior are gathered and displayed using a time-accurate  execution process. One of the distinguishing features of CodeAPeel-C is its screen that is used to animate screen programming with eleven graphics instructions. Unlike  the low-level and implicit graphics processing in real CPUs and GPUs, 
CodeAPeel-C's  pixel-based graphics instructions directly manipulate pixels in its multilayer VRAM memory, which are then projected onto its screen.  Screen programming in CodeAPeel-C is illustrated with a number of examples in Section~\ref{section:sampleRuns}. There are several auxiliary windows in CodeAPeel-C, including RAM, VRAM, STACK, and CPU Scratchpad (SPAD) registers, some of which are displayed in this snapshot. The animation extends to all the auxiliary windows with options  to change SPAD, RAM, VRAM, and STACK sizes and their bit-widths, in addition to setting byte-ordering of RAM and STACK memories to little and big endian formats, and adjusting the clock speed and other performance parameters of CodeAPeel-C processor. Programs are run as processes and several statistics are generated for each process that include CPI, IPC, total execution time in clock cycles as well as in real time. Another feature of CodeAPeel-C that sets it apart from other computer architecture simulation tools is its built-in help menu that provides extensive guidance and support for users as shown in Figure~\ref{helpWindow} that include assembly language programs for various  computer science algorithms, a description of the syntax of its instructions, and how to use its various options.
\vspace{-9pt}
\section{The Application Model}
\label{section:applicationModel}

\vspace{-7pt}\noindent
CodeAPeel-C is mainly developed using Java programming language and Java's Swing/AWT library for its user interface. For some small parts of the application, other programming languages  such as Javascript and Python are also employed. Both the user interface and core logic codebase of the CodeAPeel-C are adopted from CodeMill, which is the predecessor of CodeAPeel-C. Currently, CodeAPeel-C is formed by more than two hundred Java classes and interfaces, and developed, using object-oriented design principles. CodeAPeel-C is formed by three main components and abstractions, which are hardware modules, operating system, and user interface. It simulates six hardware components that are integrated and operated as an abstracted complete computer together. This way, the internals of an operating computer can be observed all at once by the users. The six components are divided into two types by their characteristics. The first type is related to storing and loading data, and it consists of four modules. Two of these hardware modules are RAM and VRAM, which are shared hardware modules among different processes, operating on the application. The core responsibility of RAM is to store and contain application data and object codes, which are necessary for the majority of the application and associated algorithms coded in Java. Likewise, the core responsibility of VRAM is to store the application's data and represent the layers of  CodeAPeel-C's screen, which are rendered on this screen as part of the user interface. The other two hardware modules of the first type are CPU registers and STACK, which are owned by individual processes and the data is protected implicitly from other processes. STACK is used to help execute stack instructions that provide a convenient way to compile Java programs into CodeAPeel-C's assembly and machine language programs.  Overall, CPU registers and hardware STACK behave as in most computer systems and serve as placeholders to display the results of CPU and STACK instructions of CodeAPeel-C. As such, they are used to represent and modify the application data as necessary. Additionally, these four hardware modules can be configured to different data widths and sizes. This way, different types of data representation schemes can be studied by CodeAPeel-C users. One of the most obvious benefits of configurable sizes is to be able to observe different endian storage models. The user may study little endian/big endian/no endian data store and load operations by having different data sizes for different hardware components. For example, the user may observe data store and load operations from/to 8-bit CPU registers and to/from 32-bit RAM and observe the behaviors of different endian types. These endian representations are applied to CPU registers, RAM, and STACK. Moreover, the stack growth direction may be set to ascending or descending order to observe different computer architecture models. In addition, VRAM can be configured for different resolutions and screen sizes by the user.

\vspace{-1pt}
The second type of hardware modules includes the CodeAPeel-C's CPU and Timer. These two modules  constantly work in the background, and they execute codes and trigger other modules. The timer has a relatively basic responsibility which is to signal a possible context switch for the process scheduler by a predetermined quantum time. On the other hand, the responsibility of the CPU is more central, as  a core component of CodeAPeel-C. It controls the  execution of code in both assembly and object layers. Through this process, it regulates the code within any configuration as specified by the user. These configurations may include clock rate, performance, breakpoint, and code and engine highlighting, which basically highlights any updates on the user interface to facilitate code tracking. With CodeAPeel-C's performance and clock rate options, the CPU can execute the code with the specified speed in real-time. This way, the user can both observe and feel the code in real-time, which potentially provides insight on how clock speed and performance enhancements affect the overall performance of a computer. Additionally,  users can put breakpoints on the code to debug their algorithms or observe an instant state of the computer, which is also managed by the CPU. To run all the components of CodeAPeel-C application, first the hardware components that manage data are initiated, then the timer and CPU are invoked, and then the control is turned over to the CPU. When the CPU starts running, it  retrieves the operating system at the beginning of each new instruction execution,  making CodeAPeel-C's instructions atomic in scheduling the next  process, and  the next  instruction to be executed.

\vspace{-1pt}
The operating system of CodeAPeel-C has two main responsibilities: (1) loading the programs, and (2) scheduling processes. Currently, the operating system has five programs, which are  a  compiler, parser, assembler, disassembler, and loader. The programs are generally invoked after some user settings, for example, the parser is executed on the assembly language code to check any syntax errors regarding argument formats, and truncations warnings for immediate values, when the user wants to execute an assembly language  code. Likewise, the other programs are executed whenever the user wants to compile,  assemble, disassemble, or load a code. The other important task of the operating system is  to schedule processes. When the user opens a new console in  CodeAPeel-C and conveys the execution of a code by pressing the execute button or selecting it on the commands menu, CodeAPeel-C creates a new isolated process for the code. Also, the user may initiate multiple processes. Since CodeAPeel-C can operate multiple processes simultaneously, a process scheduler is necessary. For the current version of the process scheduler, the round-robin algorithm is employed. Therefore, when the user initiates multiple processes, the processes are scheduled concurrently by context switches with intervals that are triggered by the timer. Throughout this process, the context of the processes is saved and loaded to the computer as long as the processes continue execution, and isolated executions of all processes are satisfied. During the scheduling, the processes are labeled by five different states, namely, \textit{new},  \textit{ready},  \textit{waiting},  \textit{running}, and  \textit{dead}. The process states can be changed by the user interactions using the “Step”, “Execute”, “Resume”, and “Suspend” buttons provided on the user interface. As a result, the process scheduler manages the states of processes and does necessary context switches, regarding both timer interrupts and user interactions.

All aforementioned hardware modules are included in the user interface and their states  can be monitored visually. Additionally, for all of the modules, there is an option to observe the updates in specified arithmetic settings, namely, binary, octal, signed and unsigned decimal, BCD, hexadecimal, and floating-point number (IEEE-754) representations. The user interface also provides the user with some input and output choices such as console screen, background image,  tooltips, tab view, and others. In addition, the user interface  implements the regular features of an application, such as copy and paste, undo, and redo, and all other similar features, which are found in most fully-developed applications.

\vspace{-9pt}
\section{CodeAPeel-C Assembler and Disassembler}
\label{section:assemblerDisassembler}

\vspace{-5pt}\noindent
In this section, we describe the assembler/disassembler functions, machine-level representations of instructions and their  executions in CodeAPeel-C simulator.  CodeAPeel-C provides a layer of exploration for  machine coding under its assembly language layer. Users are able to translate their assembly  and machine language programs back and forth with the help of its assembler and disassembler tools. The object  (machine) code is represented in binary format, yet users may optionally view them in hexadecimal, decimal, and octal representations as well. Moreover, they may directly manipulate object code in this layer and even write machine programs in any of the numerical representations for exploration. The object code is loaded to CodeAPeel-C's core memory, and fetched and executed on CodAPeel-C's machine processor as described in Figure~\ref{codeapeelOverview}. It can also be disassembled  back to an assembly language program and executed on CodeAPeel-C's assembly language processor. 

\begin{figure}[t]
\vspace{4pt}
\centerline{\includegraphics[scale=0.18]{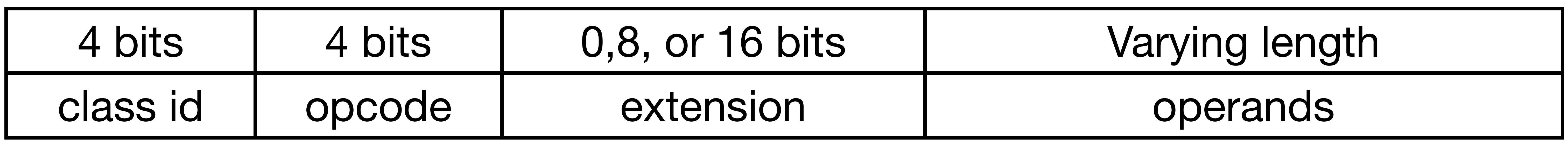}}
\vspace{-10pt}
\caption{CodeAPeel-C instruction format.}
\label{instructionFormat}
\end{figure}

Instructions of CodeAPeel-C assembly language have varying-length binary representations, that is to say each atomic instruction's length is determined by its operands. The general format of the machine level representations of instructions is shown in Figure~\ref{instructionFormat}.  Each instruction in CodeAPeel-C has a fixed size 4-bit \textit{class id} and 4-bit \textit{opcode} field. Next to the \textit{opcode} field, we have an optional \textit{extension} or \textit{function} field that can be 8 or 16 bits depending on the instruction. Remaining bits are reserved for operands of instructions whose length varies in multiples of bytes. The \textit{class id} field is used to group instructions by their functionalities. For example, instructions having functionalities related to memory reads (load instructions) have $0000$ as their class id, instructions for logical operations have $0011$ as their class id, etc. The \textit{opcode} field distinguishes the instructions within each class so that when we combine \textit{class id} and \textit{opcode} fields, we get a unique bit sequence for each distinct instruction. At the same time, instructions with parallel execution modes (vector instruction modes) may have more than one opcode. In such cases, the least significant three bits are usually the same but the most significant bit is set to $1$ to separate it from  a scalar operation.  For example, the logical OR instruction IOR has two opcodes: the opcode $0000$ represents a scalar execution mode in which CodeAPeel-C performs a logical OR operation on scalar operands and returns the result to a single destination register, but IOR instruction with opcode $1000$ represents a parallel execution mode in which CodeAPeel-C performs an OR operation over a set of source registers and  returns the results to a set of destination registers. The \textit{extension} field is used to differentiate between the number of operands in an instruction and to differentiate between register versus numerical operand choices. Instructions in CodeAPeel-C are a lot more intuitively identified than the instruction coding in RISC-V,  making instruction decoding straightforward by CodeAPeel-C's parser. This may be viewed as a trade-off between instruction format compactness and ease of decoding. As in function overloading, many instructions in CodeAPeel-C have extensions providing more depth and functionality to their operations. The extensions of instructions differ by the arity, i.e., by number of arguments that they use. Considering the IOR instruction again, it has the following four extensions:

\vspace{-2pt}
{
\begin{enumerate}
\item IOR Rx, Ry/constant/label (with two operands that perform (Rx OR Ry/constant  /label) operation and stores the result in Rx.)
\item  IOR Rx, Ry/constant/label, Rz/constant/label (with three operands that perform (Ry/constant /label OR Rz/constant/label) operation and stores the result in Rx.)
\item IOR Rx, Ry, prefix domain split, prefix direction (with four operands that prefix ORs bits of Ry and stores them to Rx. Domain split partitions the bits of Ry and specifies the domains of the prefix. Prefix proceeds from left to right when prefix direction is 0 and right to left when it is 1.) 
\item  IOR destination register set, source register set, prefix domain, prefix direction, register window (with five operands that performs a prefix OR over a set of source registers onto a set of destination registers.)
\end{enumerate}
}

\vspace{-6pt}\noindent
In these extensions, some of the operands have also varying types. For example, the second argument in extension 1 can be a register, numerical operand, or a label. To distinguish all these possible permutations of operand types, we use the extension field of the instruction format. Extensions  that are related to parallel or vector executions such as extension 4 differ also by their opcodes. Once the extension bits are set, \textit{operand} field is determined according to the operand types that are chosen. Registers are identified by an 8-bit field, providing up to 256 distinct registers,  independent of architecture configurations. On the other hand, numerical operands are set depending on whether they are immediate values or address values as well as the architecture configurations such as bit-width of registers, memory size, etc. The length of this field can be as small as zero (HLT instruction having no argument that simply terminates the execution) or indefinitely large (STI instruction that writes an indefinite size immediate value to memory at a given address).

As stated, an object code may be loaded into the memory for machine-level execution in CodeAPeel-C. Load operation is also configurable so that it provides different options to users. The user may load a program, where instructions are aligned to the memory locations according to the word sizes or the program is packed and stored as a single large string into memory. Word sizes are determined by register bit-width. When a program is loaded in word-aligned mode, no two instructions are allowed to enter the same aligned memory location, thus they are padded with zeros if necessary. The alignment options provide a means for exploration of machine-level performance as a word-aligned program would take more memory space, but decoding instructions would be simpler. On the other hand, if the program is loaded with no alignment option, less memory space would be used, but  the instruction decoding logic needs to be more complex. Endian formats are also configurable by the user for the program load operation. CodeAPeel-C also provides a core map with which users may track the programs in memory and later link them together.

\vspace{-10pt}
\section{Vectorization in CodeAPeel-C}
\label{section:vectorization}

\vspace{-8pt}\noindent
Vector processing has become popular after the introduction of SIMD instructions in Intel Pentium processors with MMX technology\cite{MMX96,millindMMX97} in mid 1990s. Others joined in, and vectorization is now part of RISC-V and other processors\cite{vectorprocessingAsanovic,vectorization,armVector,asanovicDiss}. Essentially, vectorization amounts to performing a given instruction over a vector or vectors of operands. In CodeAPeel-C, scalar registers constitute the individual operands in vectors. For example, in a 16-register SPAD, two vectors of eight operands  may be created (initialized) by two immediate vector load instructions and then added by a vector add instruction.  All register instructions have been overloaded in CodeAPeel-C, and one of the   instruction formats is designed to implement vector operations. Depending on the type of instruction, one or two fields are set aside to identify vectorized operands. Unlike in real processors, vectors need not be made up of consecutive registers in CodeAPeel-C. Instead, registers that form a vector are identified by binary strings that may be viewed as masks. For example, INC 5,XAF,0; increments each of the registers R0, R2, R4, R5, R6, R7 by 5 in a SPAD with eight registers as dictated by the 8-bit mask 10101111. Instruction formats for binary vector instructions are more involved, and they include prefix operations, where ordinary vector operations become special cases of such prefix operations. For example, the following CodeAPeel-C program 

\vspace{-8pt}
{\small
\begin{verbatim}
        LDI R0, 3; LDI R1, 7; LDI R3, 35; ADD X07,XD0,0,0,0; 
        //prefix domain-width = 0, prefix direction: 0, prefix window = 0
\end{verbatim}
}

\vspace{-8pt}\noindent
computes the prefix sums: R5 = R0 = 3; R6 = R1 + R0 = 10;  R7 = R3 + R1 + R0 = 45, where the masks X07 = 00000111 and XD0 = 11010000 select registers R0, R1, and R3 as the operands in the source vector, and R5, R6, and R7 as results holders  in the destination vector. The prefix sums are then computed over the source vector and stored into the destination vector.

\noindent
The three parameters, {\em prefix domain}, {\em prefix direction} and {\em prefix window} determine (i) the domain of registers within an SPAD to which the prefix operations are applied, (ii) the direction of the prefix, i.e., left to right or right to left, and (iii) the SPAD window in which the register operands are located.  As another example, consider adding two 4-dimensional vectors in CodeAPeel-C. The following CodeAPeel-C program illustrates how this is done. 

\vspace{-4pt}
{\small
\begin{verbatim}
LDI R0, 27; LDI R1, 10; LDI R2, 6;   LDI R3, 3; 
LDI R4, 10; LDI R5, 20; LDI R6, 30; LDI R7, 40;
ADD XFF,XFF,2,1,0; 
//domain-width = 2, prefix direction: 1, prefix window = 0
//Domain is divided into four subdomains: {R0,R1},{R2,R3},{R4,R5},{R6,R7}
//Prefixes for the four subdomains are 
//R1 = R1 = 10, R0 = R0 + R1 = 37, R3 = R3 =  3, R2 = R2 + R3 = 9,
//R5 = R5 = 20, R4 = R4 + R5 = 30, R7 = R7 = 40, R6 = R6 + R7 = 70. 
\end{verbatim}
}

\vspace{-4pt}\noindent
The register additions may be viewed as adding two vectors (27,6,10,30) and (10,3,20,40) and storing the resultant vector into R0, R2, R4, and  R6. The same vectorization concept is applied to other binary arithmetic and logic instructions much the same way. Register  shifts are handled slightly differently, where prefixing is interpreted as a concatenation of registers, where shift and move instructions become special cases as well. Vector moves or copies are also more directly implemented using binary incidence matrices as masks. For example,

\vspace{-6pt}
{\small 
\begin{verbatim}
              LDI R0,1; LDI R1,2; LDI R2,3; LDI R3,4;
              LDI R4,5; LDI R5,6; LDI R6,7; LDI R7,8;
              MOV 2,8,4,1,0;
\end{verbatim}
}
\vspace{-4pt}\noindent
Expanding 2,8,4,1 as a binary sequence gives 

{\small
\begin{center}
\vspace{-2pt}\noindent
0 0 0 0 0 0 1 0,   0 0 0 0 1 0 0 0,   

\noindent
0 0 0 0 0 1 0 0,    0 0 0 0 0 0 0 1.

\vspace{-8pt}
\end{center}
}
\noindent
or in matrix form, it becomes:

\vspace{-7pt}
{\small \begin{verbatim}
                    0 0 0 0 0 0 1 0 R0 = R6 = 7
                    0 0 0 0 1 0 0 0 R1 = R4 = 5
                    0 0 0 0 0 1 0 0 R2 = R5 = 6
                    0 0 0 0 0 0 0 1 R3 = R7 = 8
\end{verbatim}
}

\vspace{-6pt}\noindent
The rows of the matrix represent registers R0, R1, R2, R3, whereas the columns represent registers R0, R1, R2, R3, R4, R5, R6, R7. The '1' entries represent the copies from R6 to R0 (1st row), R4 to R1 (2nd row), R5 to R2 (3rd row), and R7 to R3 (last row). Thus, the vector move in this example amounts to replacing vector (R0,R1,R2,R3) by vector (R6,R4,R5,R7) with four concurrent moves. The last argument in the MOV instruction, i.e., 0 indicates that MOV must be applied to first eight registers. If the same argument is 1 then the window shifts to the next group of eight registers, and the vector MOV instruction amounts to executing 
MOV R8,R14; MOV R9,R12;  MOV R10,R13;  MOV R11,R15; all in parallel.

Vector instructions in CodeAPeel-C are natural extensions of scalar instructions together with prefix operations in arithmetic instructions,  parallel moves in the case of move instructions, and simple replication of a scalar operation on all selected registers in unary instructions such as INC, DEC, CMP, etc. In other extensions such as vector shuffles and swaps, the vectorization occurs by applying shuffles and swaps to individual registers with shuffle and swap widths that are specified in the instruction. In  register shifts and rotations, and floating-point instructions, vectorization facilitates floating-point operations on multiple registers together with concatenation of registers to extend the length of  operand bit strings, and precision and range of floating-point operands. 

\vspace{-8pt}
\section{Screen Programming in CodeAPeel-C}
\label{section:screenProgramming}

\begin{figure}[t]
\vspace{2pt}
\centerline{\includegraphics[scale=0.36]{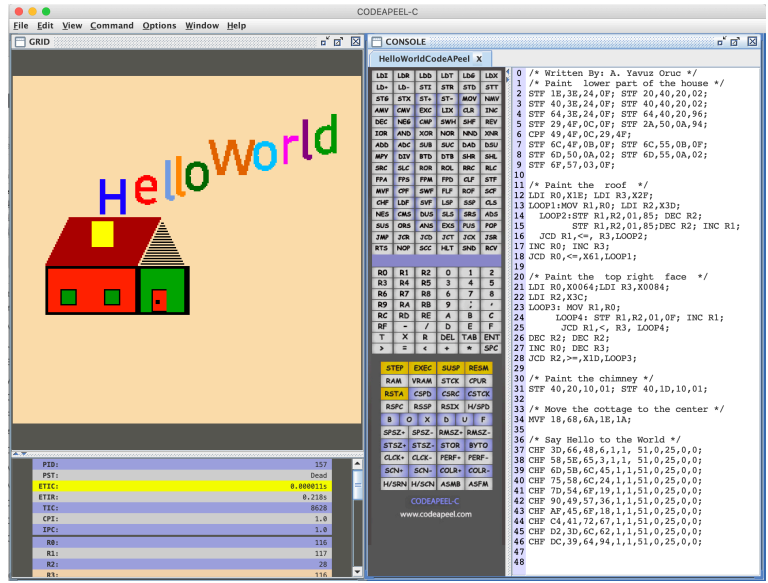}}
\vspace{-6pt}
\caption{An example of screen programming in CodeAPeel-C.}
\label{helloWorld}
\end{figure}

\vspace{-5pt}\noindent
Pixel-based screen programming is unique to CodeAPeel-C among all  computer architecture educational tools. An 8-layer,  pixel graphics screen is embedded into its architecture,  where pixels are manipulated with eleven graphics instructions to create, clear, copy, move, swap, rotate, flip, scale,  load and save pixelated images, and  to display characters as illustrated in Figure~\ref{helloWorld} with a visual ``Hello World'' program.  The multilayer graphics provide CodeAPeel-C the capability to produce pixel-based animations with just a few screen instructions and simple loops.
All of these screen instructions are implemented  in Java by rendering a combination of predefined images and multiple layers of dynamically updated buffered images. When a screen instruction is executed, both CodeAPeel-C's  Video RAM (VRAM) and its screen are updated in real time. Each buffered image represents a layer on CodeAPeel-C's screen. The number of layers, pixel resolution of the screen, screen width and height, and visible view are all reconfigurable. In addition, if a desirable virtual  CodeAPeel-C screen size  does not fit into the width and/or height of the window that contains the screen, then the window scrolls are activated automatically to allow resizing with a scrollable screen view. Screen resolution is adjusted by fixing the pixel size, which provides zoom in/out functionality on CodeAPeel-C screen. Virtual pixel size may also be used to simulate different screen resolutions. CodeAPeel-C screen instructions also provide  image copying  between VRAM and RAM in order to create images before they are displayed, or to save them after they are displayed.  All screen instructions are overloaded incrementally in their syntax  to support  manipulation of pixel frames with different shapes such as square, rectangle, polygon, etc. For example, the Set Frame (STF) instruction has the following five forms with an increasing number of fields to draw more complex shapes:

\vspace{-4pt}
{
\begin{enumerate}
\item STF frame x-address, frame y-address, frame-width, frame-color;  Draws a square in layer 0;
\item STF frame-x address, frame-y address, frame-width, frame-color, layerN;  Draws a square in layerN;
\item STF frame-x address, frame-y address, frame-width, frame-height, frame-color, layerN; Draws a rectangle in layerN;
\item STF frame x-address, frame-y address, frame-width, frame-height, frame angle, interior color, layerN; Draws a rotated rectangle   with a specified angle in  layerN; 
\item STF frame address-x, frame address-y, edge-count, edge-width, angle, interior color, border color, layerN; Draws a polygon with a specified number of edges, edge-width, angle, interior and border colors in layerN; 
\end{enumerate}
}

\vspace{-4pt}\noindent
Other screen instructions are similarly overloaded to clear, move, copy, swap, flip, rotate, scale, load, and save different types of frames, and display characters in various fonts, and font  styles and sizes on CodeAPeel-C screen.

\vspace{-12pt}
\section{Sample Runs}
\label{section:sampleRuns}

\vspace{-6pt}\noindent
In this section, we provide three examples to illustrate the utility of CodeAPeel-C as a multilayer computer architecture  teaching and learning  tool. As we described in the earlier section, one of the key features of CodeAPeel-C as an instructional tool is the simulation of a computer screen. Its eleven graphics instructions and 8-layer graphics provide an ideal tool to integrate assembly and machine layer programs with pixel processing and screen programming. This provides the user with the capability to seamlessly mix images and data  directly in the multiple layers of CodeAPeel-C as the following examples illustrate.

\vspace{-12pt}
\subsection{\rm\em Binary-Tree Search in Assembly-language Layer}

\begin{figure}[t]
\vspace{6pt}
\centerline{\includegraphics[scale=0.37]{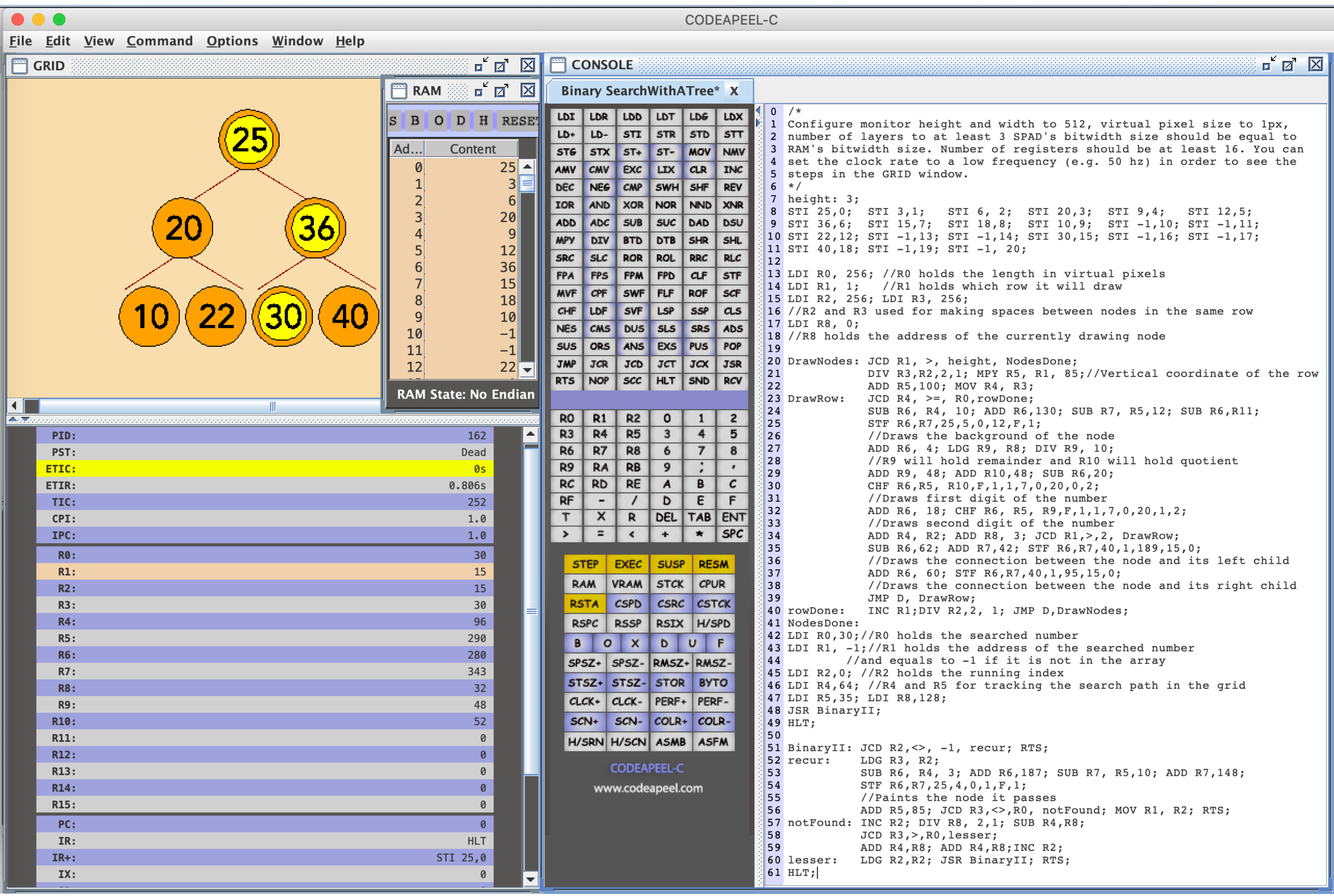}}
\vspace{-3pt}
\caption{Binary tree search in assembly layer.}
\label{binaryTreeSearch}
\end{figure}

\vspace{-6pt}\noindent
The binary-tree search is a basic computer algorithm that is frequently covered in introductory computer science and engineering courses. A visual implementation of binary search in CodeAPeel-C involves three interconnected tasks: (1) create a binary tree on CodeAPeel-C's screen, (2) search the binary tree, (3) animate the search process on the screen. The first task is carried out using the STF (Set Frame) and CHF (Character Frame) instructions of CodeAPeel-C. The search and  search animation tasks are implemented as a simple recursive function using CodeAPeel-C's JSR (Jump to Subroutine) and RTS (Return from Subroutine) instructions. These instructions work with CodeAPeel-C's STACK to make subroutine calls an easy process. A snapshot of  the CodeAPeel-C session performing these three steps is shown in Figure~\ref{binaryTreeSearch}, where 30 is searched. The path of the search is animated while the code is executed and animated in slow motion using the STF (Set Frame) and CHF (Character Frame) screen instruction together with other instructions, and by adjusting the clock rate from the options menu for the user to observe how the process of binary search proceeds.

\vspace{-12pt}
\subsection{\rm\em Towers of Hanoi in Machine Layer}

\begin{figure}[t]
\vspace{10pt}
\centerline{\includegraphics[scale=0.45]{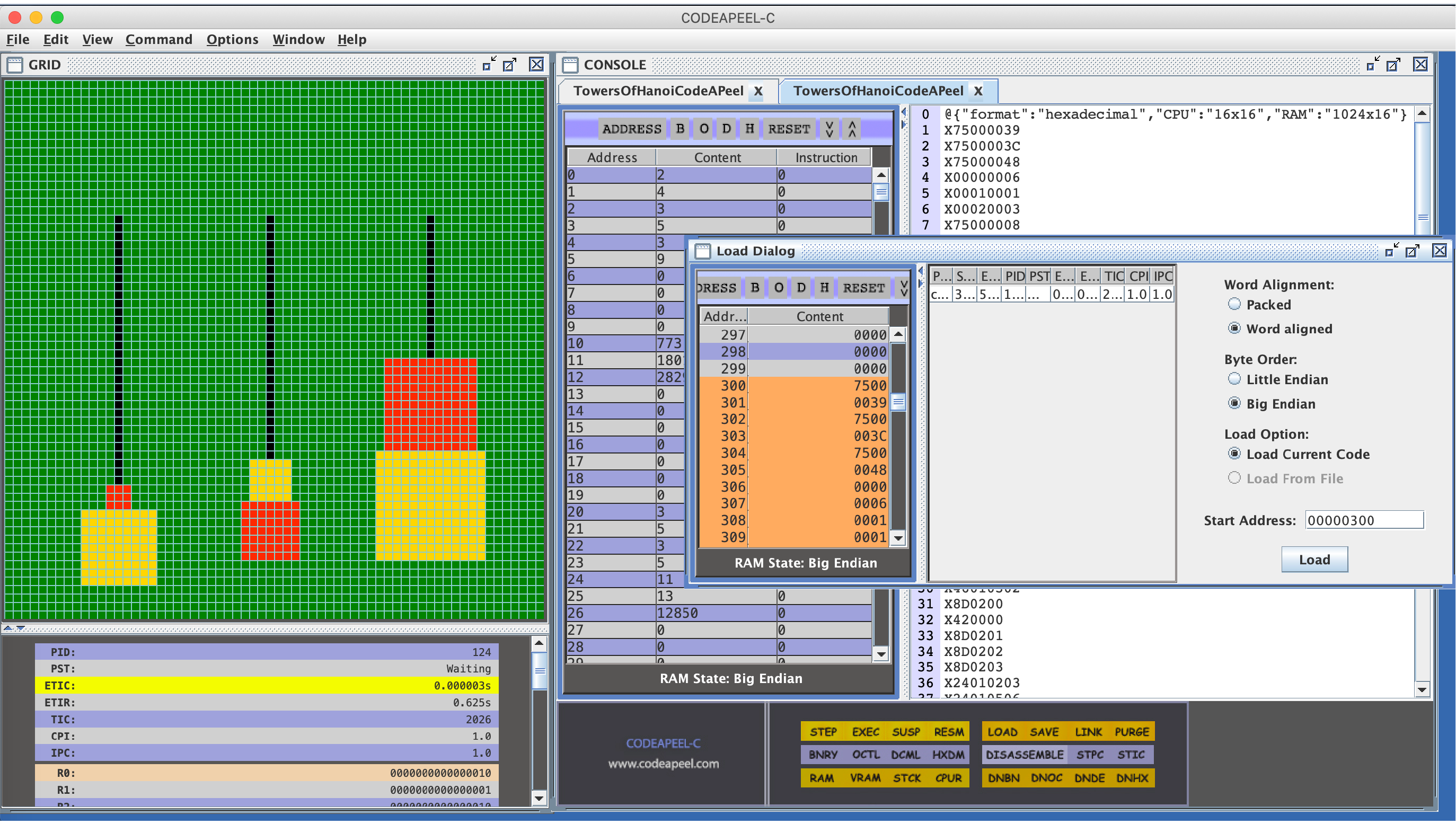}}
\vspace{-2pt}
\caption{Implementing Towers of Hanoi in machine layer of CodeAPeel-C.}
\label{towersOfHanoi}
\end{figure}

\vspace{-6pt}\noindent
Our second example illustrates how Towers of Hanoi, another widely-used algorithm  that is used to describe the power of recursion, is  implemented visually in CodeAPeel-C using its screen. Here, we have several subtasks the user may use CodeAPeel--C to explore, such as creating the stacks (pegs) and disks, placing them on CodeAPeel-C's screen, and writing a recursive function to move the disks in the right order between the pegs. As in the earlier example, these subtasks can easily be accomplished using CodeAPeel-C instruction repertoire, its CPU registers, STACK, RAM, machine layer view of the program window and the loader as seen in Figure~\ref{towersOfHanoi}. While animating the stacks moving between the pegs on the screen, the user can view how various operands in STACK, RAM, CPUR are updated, and instructions in machine code of Towers of Hanoi code in the program window are executed. In addition, the clock rate can be adjusted to speed up and slow down the animation process. 

\vspace{-12pt}
\subsection{\rm\em Evaluation of BNF Expressions}

\begin{figure}[t]
\vspace{10pt}
\centerline{\includegraphics[scale=0.42]{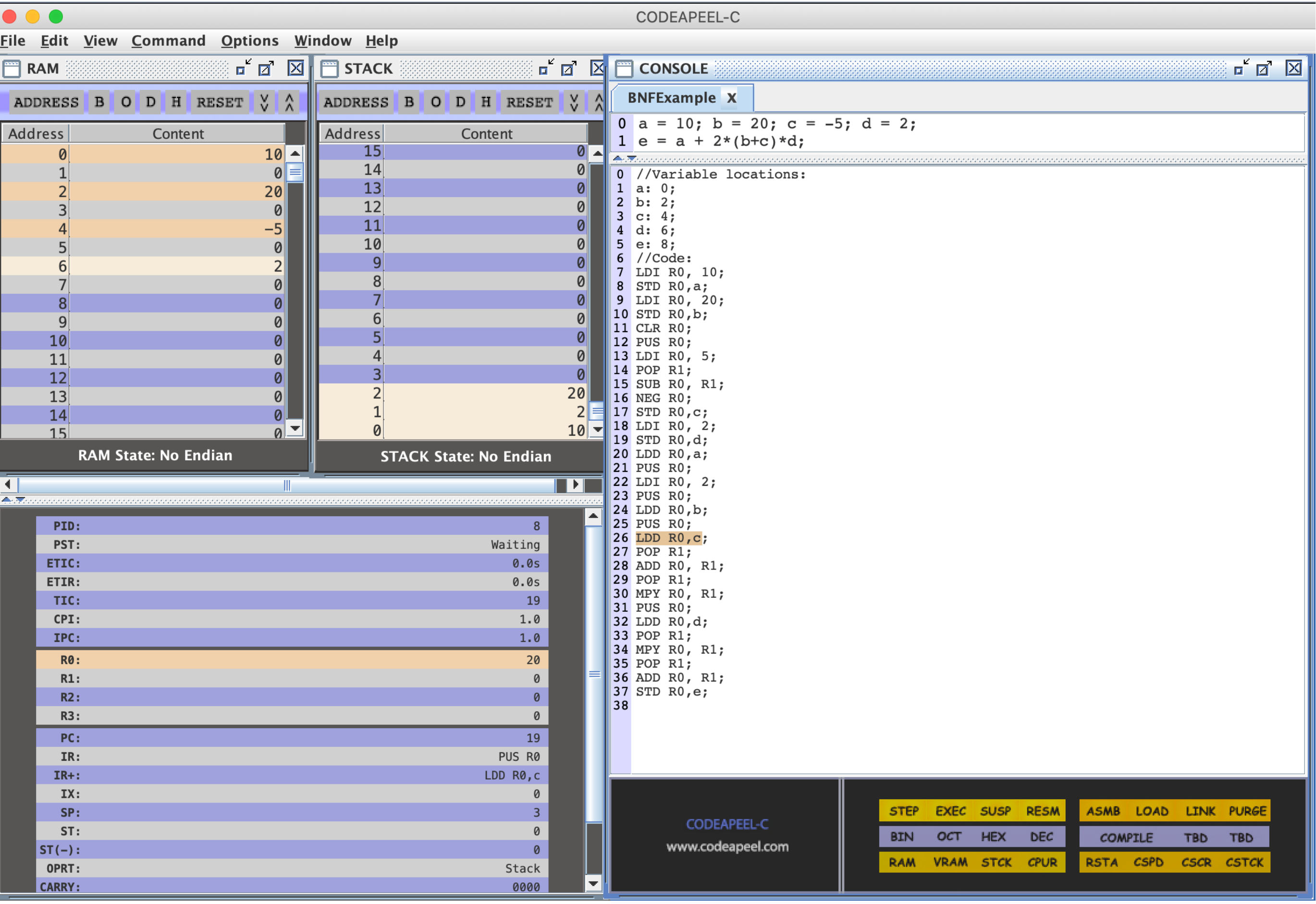}}
\vspace{-2pt}
\caption{A BNF compilation in CodeAPeel-C.}
\label{BNF}
\end{figure}

\vspace{-6pt}\noindent
Backus-Naur-Form (BNF) expressions describe context-free grammars and help define the syntax of  high-level programming languages. CodeAPeel-C support translations of BNF expressions into assembly and machine language programs, consisting of instructions in its instruction set repertoire. The example in Figure~\ref{BNF} shows the compilation of the following BNF program into CodeAPeel-C assembly language program.
\vspace{-3pt}
\begin{verbatim}
                   a = 10; b = 20; c = -5; d = 2;
                   e = a + 2*(b + c)*d;
\end{verbatim}

\vspace{-3pt}\noindent
The translation process makes use of CodeAPeel-C's RAM and STACK in order to generate the corresponding assembly language program.  CodeAPeel-C also supports compilation of simple Java programs into CodeAPeel-C assembly and machine language programs. Future editions of CodeAPeel-C will support compilation of C/C++, and other high-level programming languages.

\noindent
These examples  demonstrate  only a few of the salient features of CodeAPeel-C as an instructional tool for hands-on teaching and learning of fundamental concepts in computer architecture courses. The options menu has many settings that extend from selection of number representations,  to setting byte-ordering to big endian versus little endian or no endian, assigning CPI values to CodeAPeel-C instructions, turning on and off  the soft warnings and  tooltips,  highlighting during the execution of programs, and  the truncation of operands, etc. All of these options are designed and built into CodeAPeel-C to explore the performance of  assembly and machine language programs as they run   over its instruction set architecture and simulation environment. The help window provides many other examples to guide the user to take advantage of many other features of CodeAPeel-C in order to create more complex examples for a more in-depth understanding of the core computer architecture concepts. 

\vspace{-10pt}
\section{CodeAPeel-C v.s Other Tools}
\label{section:otherTools}

\vspace{-7pt}\noindent
It will be instructive to compare CodeAPeel-C with other educational computer architecture tools to highlight its key features. Such a comparison is provided in Table~\ref{comparisonTableL}. MARS does not a  have built-in pipelining simulation, but a plug-in for it is reported in \cite{MarsPlug-in}. All  computer architecture education tools  in the table, other than CodeAPeel-C are designed for simulating MIPS or RISC-V instruction set architecture, where some  focus on execution of machine programs, and some provide pipelining diagrams to demonstrate how MIPS/RISC-V pipelining works as seen in the table. In contrast, CodeAPeel-C is designed and developed to provide a baseline  architecture that combines scalar and vector instructions together without being tied up to any particular architecture. We expect to target open source architectures such as RISC-V and other real processors in future editions of CodeAPeel-C as described in Figure~\ref{codeapeelOverview}.  Even though the current version of CodeAPeel-C does not support pipelining and caching, future versions are expected to also include  both these features as the main goal of CodeAPeel-C is to integrate all layers of computer architecture into a single application without any plug-ins and too many disconnected windows.

\begin{table}[t]
\vspace{8pt}
\centerline{\includegraphics[scale=0.39]{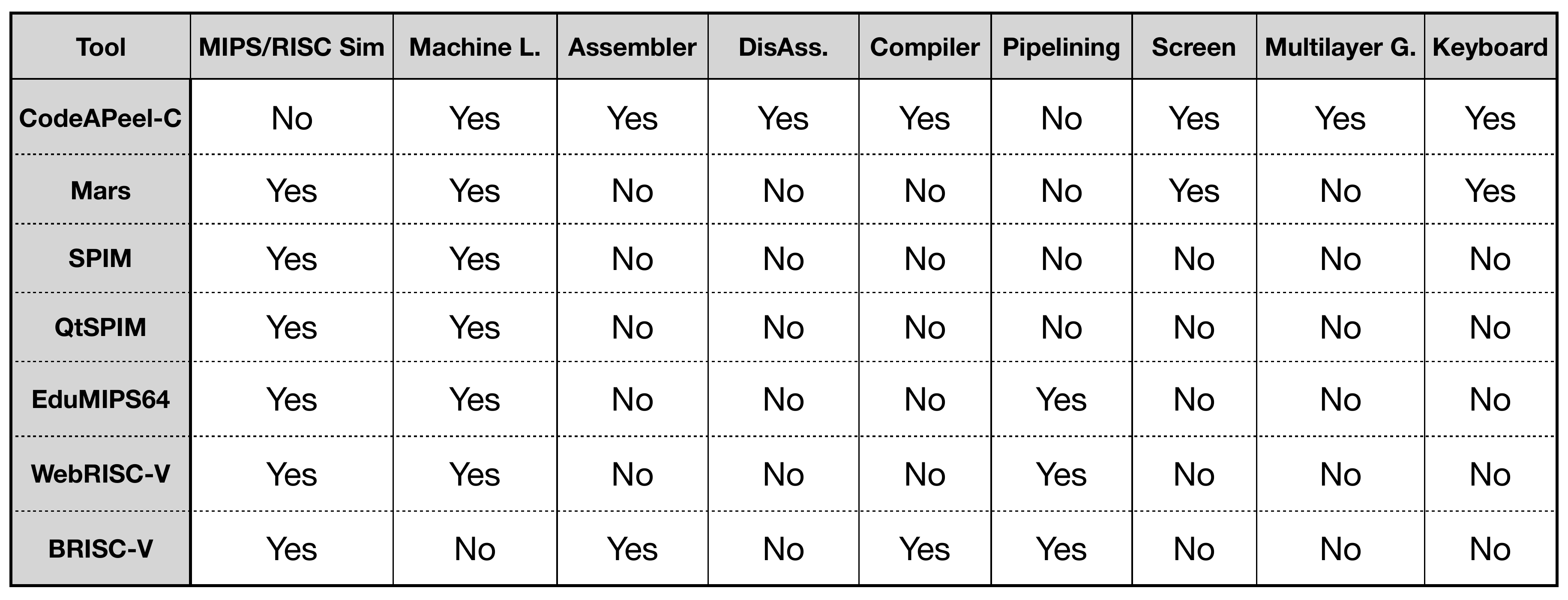}}
\vspace{4pt}
\caption{ CodeAPeel-C versus other simulation tools.}
\label{comparisonTableL}
\end{table}

\vspace{-12pt}
\section{Conclusions and Future Work}
\label{section:conclusion}

\vspace{-8pt}\noindent
We presented CodeAPeel-C, an integrated, multilayer teaching and learning tool for computer architecture courses, and  described its design and development in Java. 
The current version of CodeAPeel-C has its own generic RISC instruction set architecture that supports all kinds of instructions, which are found in real RISC processor chips, in addition to screen instructions that make it a unique learning tool with an integrated computer monitor for visualization of computer algorithms during their assembly language and machine layer executions. Vector mode in CodeAPeel-C adds another dimension to it, and makes it a dual-instruction set architecture that support both SISD and SIMD instructions. A  multiple machine-target feature will be added to the next version of CodeAPeel-C to make it a more universal learning tool that can be used to run assembly and machine programs on real instruction set architectures such as ALPHA, ARM, MIPS, SPARC, RISC-V,  and others.  In the  longer time frame, we anticipate adding pipelining and cache components as well as a microprogramming layer to CodeAPeel-C to make it a more comprehensive learning tool for computer science and architecture courses.

\vspace{4pt}\noindent
{\bf Acknowledgements:} CodeAPeel-C (Originally CodeMill) was conceived by the first author and co-developed by him,  Emre Gunduzhan, Hidayet Aksu, Abdullah Atmaca, Yusuf Nevzat Sengun, and Ali Semi Yenimol. The user interface was originally created in a much simpler form in CodeMill, the forerunner of CodeAPeel-C. The coding contributions of Gonca Yılmaz, Ezgi Yavuz, and Ahmet Berk Eren to CodeAPeel-C are gratefully acknowledged. Funding for CodeAPeel project is provided by AlgoritiX,  a small business company, which is  privately owned by A. Yavuz Oruc, Cagdas Dirik, and Abdullah Atmaca.

\noindent

\vspace{-15pt}

\end{document}